\documentclass[envcountsame]{llncs}

\usepackage{graphicx}
\usepackage{amsmath}
\usepackage{color} 

\newcommand{\tmbraw}[3]{(\ensuremath{#1}, \ensuremath{#2}, \texttt{#3})}
\newcommand{\tmb}[3]{\tmbraw{#1}{#2}{#3}}
\newcommand{\token}[2]{\texttt{#1}$^{#2}$}

\begin{document}

\title{Anonymous On-line Communication Between Program Analyses}
\subtitle{(Regular paper)}
\author{Marek Trt\'{i}k}
\institute{
    \email{trtikm@gmail.com}
    }

\maketitle

\vspace{-0.25cm}
\begin{abstract}
We propose a light-weight client-server model of communication between program
analyses. Clients are individual analyses and the server mediates their
communication. A client cannot see properties of any other and the communication
is anonymous. There is no central algorithm standing above clients which would
tell them when to communicate what information. Clients communicate with others
spontaneously, according to their actual personal needs. The model is based on
our observation that a piece of information provided to an analysis at a right
place may (substantially) improve its result. We evaluated the proposed
communication model for all possible combinations of three clients on more than
400 benchmarks and the results show that the communication model performs well
in practice.

\end{abstract}

\section{Introduction}
\label{sec:Introduction}

The most common way how researchers combine program analyses starts by selecting
two or more existing analyses and then follows an invention of an algorithm
combining them into a new analysis. This process has two obvious issues. First,
the selection of analyses is often based only on our intuition, because we have
no evidence about dispositions of individual analyses for their mutual
cooperation. Indeed, this is actually what we are about to discover. If we
further realise that for $ n $ existing analyses we can create $ 2^n - n - 1 $
combined analyses (where some combinations may be less efficient than others),
then our intuition may easily lead as to a less promising choice. The second
issues is the complexity in the process of inventing of the new algorithm. The
process is so complex because we do not have solid data we could analyse in
order to see promising directions in our research. We can basically rely on our
intuition again.

In this paper we propose an alternative approach in which analysers exchange
information by anonymous communication. An analyser asks others for an
information according its actual personal needs. There is no central algorithm
which would tell analysers when to communicate what information.

The approach is heavily based on an observation which we can
experience during experimentation with many existing analysers:
\emph{A~piece of information provided to an analyser at a right place may
(substantially) improve its result}.

By a \emph{right place} we mean any line in the source code of an analyser where
a loss of precision may occur or where an additional information may simply
help. For example, a path insensitive analysis typically loses the precision in
join nodes of a control flow graph. So, all lines in the source code of the
analyser where the join operation is invoked are right places. Similarly,
widening operator of an abstract interpreter typically causes a loss of
precision. So, each line where the operator is called is the right place.
Further, many analyses over- or under-approximate semantics of some operators,
like bit-operators (e.g. \texttt{\&}, \texttt{|}, \texttt{\^}, in \texttt{C}),
floating point arithmetic, operators of fixed size integers (i.e.~ignoring over-
or under-flows), pointer arithmetic operators, etc. All places in the source
code where these over- or under-approximated operators are invoked are right
places. Clearly, it should not be difficult for a researcher to identify right
places in his analyser, because he should be familiar with the source code.

The goal of the approach is to allow a delivery of information to right places
identified in an analyser (to allow reduction of the precision loss occurring
there). So, we extend the code in each right place by issuing communication
queries to other analysers. For example, a query may ask other analysers for
possible values of a certain variable at a given control location. We also need
to extend each right place by a code translating the received information into
data which will be then used in that right place as usual (i.e.~as any other available data).

An analyser is further supposed to answer queries from other analysers. For
example, a received query for possible values of a certain variable at some
control location may involve a search for the variable in analyser's memory
model and encoding the retrieved values into a response message. Code for
answering queries can be added into an analyser as a separate \emph{answering
module}.

If an analyser uses an information from other analysers in some of its right
places, then we can ask whether correctness and termination of its analysis is
still preserved or not. For example, an analyser may fail to build program's
over-approximation if its computation is affected by incomplete information from
other analysers who build program's under-approximation. Therefore, an analyser
may also need a code issuing \emph{special queries}, which allow it to ensure
correctness and termination of whole its extended source code \emph{without
consideration of properties of any other analyser}.

Once we extended $ n $ analysers $ A_1, \ldots, A_n $ as was outlined above, we
can use them for simultaneous and cooperative analysis of a given program:
\begin{center}
\includegraphics[scale=1]{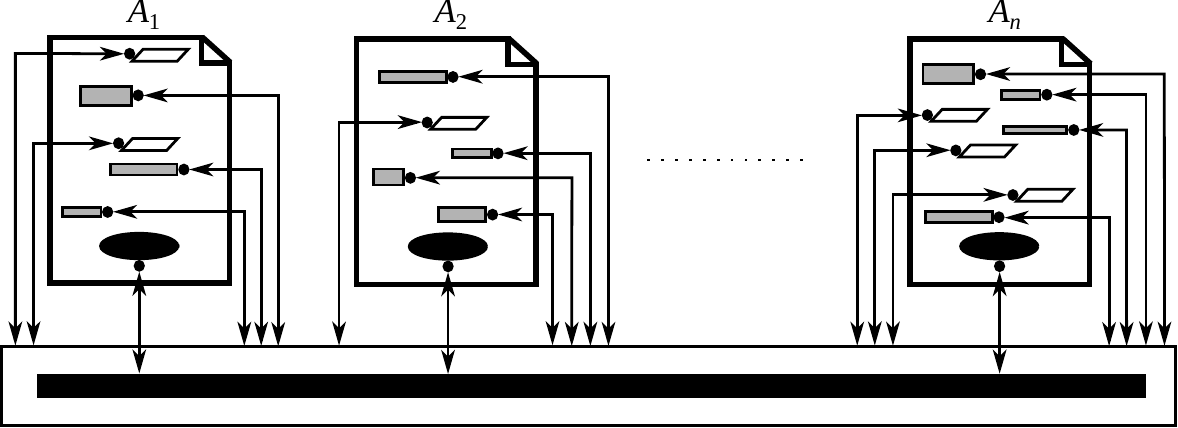}
\end{center}
For each analyser $ A_i $ we see its source code
\includegraphics[scale=1]{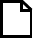} with
extended right places
\includegraphics[scale=1]{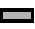}, places
with special queries
\includegraphics[scale=1]{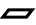}, and the module
for answering queries
\includegraphics[scale=1]{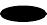}. Arrows
show the communication channels (information flow). All communication is
mediated through a shared medium
\includegraphics[scale=1]{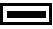}. The
communication is thus in the client-server style. Clients are individual
analysers and the medium stands for the server. Observe that each analyser
communicates only in fixed and a priori determined places in its code: (1) in
the extended right places, (2) in the places for special queries, and (3) in the
module for answering queries. All other code is intact and operates as usual.

A communication query of any kind can only be issued from a client (when its
execution reaches
\includegraphics[scale=1]{communication_scheme_right_place} or
\includegraphics[scale=1]{communication_scheme_special} place). The
query then goes directly to the server. This is depicted in the scheme by arrows
from \includegraphics[scale=1]{communication_scheme_right_place} or
\includegraphics[scale=1]{communication_scheme_special} to the outer
box \includegraphics[scale=1]{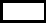} of
the server. The server then broadcasts the received query trough the channel
\includegraphics[scale=1]{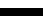} to
response modules
\includegraphics[scale=1]{communication_scheme_response_module} of all
clients (including the one initiating the query). Once the server receives
answers from response modules of all clients it combines them into a single
response message and sends it back to the place
\includegraphics[scale=1]{communication_scheme_right_place} or
\includegraphics[scale=1]{communication_scheme_special} where the query
was originally issued.

The biggest challenge in this approach is how to deal with incompatibility of
individual analysers. First of all, each analyser builds its own internal
representation of an analysed program. For example,
CPAchecker~\cite{CPAcheckerURL} builds a control flow graph (CFG) directly from
a \texttt{C} source code such that edges are labeled by corresponding \texttt{C}
expressions. Bugst~\cite{BugstURL}, in contrary, translates the \texttt{C}
program into LLVM and the CFG is constructed from the assembly.
KLEE~\cite{KLEEURL} goes even further, because it applies several compiler
optimisations to the LLVM translation. Control flow graphs constructed by
analysers for the same program thus almost always differ in numbers of control
location, branchings, join nodes, composition of loops (e.g.~due to
optimisations), function inlining, and so on. In this setting a query ``give me
possible values of the variable \texttt{a} at the control location 10'' can be
clear only for the analyser who issues this query. Indeed, another analyser may
have quite different interpretation for the ``location 10'' and the variable
\texttt{a} may be unaccessible there. Moreover, the other analyser may use
different names for variables (e.g. due to transformation to LLVM), so the
``variable \texttt{a}'' can be completely unknown identifier.

The second issue is that each analyser builds its own model of program's memory.
For example, stack variables can be organised differently in different tools, so
their addressing may also differ (note that we cannot rely on names of
variables). Dealing with program's heap is especially difficult, because
representation of address space may differ significantly. For example, some
analysers assign unique identifiers to newly allocated blocks, some analysers
recycle the identifiers when blocks are released, other do not, and
dynamic analyses use physical addresses (e.g. in testing).

We resolve the incompatibility of internal program representations by
introducing unifying representation called \emph{canonical program}, and we
resolve the incompatibility in models of address spaces by introducing unifying
representation called \emph{canonical memory}. These canonical representations
do \emph{not} replace original internal representations in analysers. Each
analyser still performs its analysis on its own internal representations. Both
canonical representations are used exclusively for the purpose of communication.
Namely, each communication query starts with its translation from terms of
analyser's internal representations to terms of canonical representations, and
once the query is received by an analyser, the first thing is its opposite
translation. In the scheme above there are depicted all these \emph{translations
places} using the symbol
\includegraphics[scale=1]{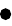}. We
see that data in canonical representations flow only through the server.

Now we have a general overview of the approach. It remains to define precisely
what queries may clients actually issue to the server, what are their exact
properties and requirements, and how both canonical representations look like.
All these things together represent a \emph{communication protocol} of our
approach. We define it in Section~\ref{sec:CommunicationProtocol}. Namely, in
Section~\ref{sec:CommunicationQueries} we describe what communication queries we
need and why, then in
Sections~\ref{sec:CanonicalProgram}--\ref{sec:CanonicalMemory} we define both
canonical representations, and in Sections~\ref{sec:Client} and \ref{sec:Server}
we formally define all queries for clients and the server using the introduced
terminology.

As we already mentioned, the approach does not allow a client to insert a code
which would be somehow related or dependent on any other client. This
requirement is necessary, because communication of client during an analysis of
a program is completely anonymous. The positive side of that requirement is that
once the communication protocol is integrated into a client, the client can be
run with any other clients (also implementing the protocol). On the other hand,
clients exchange information in rather unorganised manner. Indeed, a client asks
for an information whenever its execution reaches some of its right places. We
have thus very right to ask how much effective this unsupervised data exchange
can be in practice. In this context we emphasise the importance of
Section~\ref{sec:Evaluation} where we present results from our experimental
evaluation.

\section{The communication protocol}
\label{sec:CommunicationProtocol}

A \emph{client} is a program analyser or a program analysis inside an analyser
which is able to communicate with other clients during analysis of a given
program. A \emph{server} is a program utility (or a module) mediating the
communication between clients. A client can only communicate with the server and
has no information about other clients, except their count. Data exchanged
between the server and a client are received in exactly the same order as they
are sent.

There is a \emph{time-out} for the whole communication common to all clients and
the server. After the time-out both the server and any client may completely
ignore any communication queries. Also, any communication query not terminated
before the time-out can be terminated immediately without any response.

Given a program written in a certain programming language, a \emph{concrete
program state} is  an element of the concrete semantics of the language, an
\emph{abstract state space} is any subset of a client's interpretation
(e.g.~abstraction or generalisation) of the semantics of the language, and an
\emph{abstract program state} is an element of an abstract state space.

\subsection{What communication queries we need?}
\label{sec:CommunicationQueries}

The purpose of this section is to explain on an intuitive level what
communication queries we established for the protocol and why. We shall survey
principles of the communication from the point of view of a single client. We
will observe the client in different situations from which purposes of
individual communication queries will become apparent.

We suppose our client performs the standard interval
analysis~\cite{IntervalAnalysis}. And for simplicity of the presentation we put
the client into an ideal setting: All communicating clients use the canonical
program and memory as their internal representations and they use same variable
names for same addresses.

Let clients analyse the following simple program (nodes model the instruction
counter, solid edges represent program transitions, and dotted edges represent
arbitrary number of (unimportant) solid edges):
\begin{center}
\includegraphics[scale=0.9]{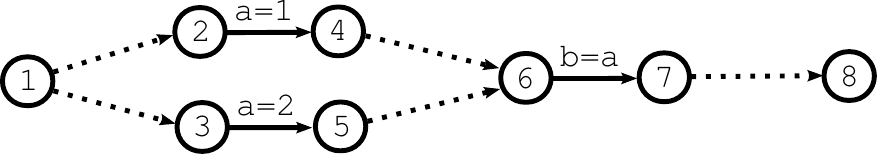}
\end{center}
We start the observation in the moment when
our client already performed its analysis along paths 1,2,4,6 and
1,3,5,6. So, value of \texttt{a} at the node 6 is the interval [1,2]. Since the
node 6 is a join node (a loss of precision may occur there),
the client decides to get value of \texttt{a} also from other
clients, in order to compute a more precise value for \texttt{b} at the node 7.
Therefore, the client issues ``\emph{get values}'' query to all clients (through
the server) at the node 6 for values of \texttt{a} computed for both paths
leading to the node 6. Each client is supposed to return an
\emph{over-approximation} of its \emph{current} knowledge about the queried
memory. And the server returns to the client an \emph{intersection} of
individual over-approximations.

Let us suppose one of the clients is ``evil''. It explored only the path 1,2,4,6
so far, so it only knows that the variable \texttt{a} has value 1 at the node 6.
It thus returns the equality $ a = 1 $ as an over-approximation of its current
knowledge about \texttt{a}. Note that it could also return formulae like $ a > 0
$, $ a = 1 \vee a > 5 $, or even \emph{true} (i.e.~any possible value for
\texttt{a}) as an over-approximation. Of course, clients attempt to return the
most precise over-approximations. As a consequence our client receives from the
server a \emph{conjunction} $ a = 1 \wedge (1 \leq a \leq 2) \wedge \varphi_1
\wedge \ldots \wedge \varphi_n $, where $ a = 1 $ is the answer from the evil
client, $ 1 \leq a \leq 2 $ is the answer from our client (a representation of
the interval [1,2]), and $ \varphi_i $ are answers from all other clients. Let
us suppose the formula is equivalent to $ a = 1 $. Then our client computes the
value of \texttt{b} at the node 7 as the interval [1,1]. If this value is not
later recomputed to the correct over-approximation [1,2], our client may finish
the analysis with an unsound result, i.e.~without achieving of an
over-approximation of the program's memory.

In order to prevent such situations each client \emph{has to} notify all other
clients (through the server) about \emph{each} change in its abstract state
space. For example, when the evil client updates its abstract state by executing
the edge (3,5), it has to send a notification ``\emph{on location outdated}'' to
all clients. The notification is coupled with the node 5 and the path 1,3,5 to
describe where the change was made and what program paths were considered in the
update. Other clients react to the notification s.t. they recompute the
corresponding parts of their state space (this may trigger re-computations of
transitively dependent parts). Note that it is sufficient for a client to react
only to those notifications whose nodes and paths inform the client that some of
its ``\emph{get values}'' queries issued so far may now return different
answers.

Let us return to the scenario where the computation of our client was based
on the answer $ a = 1 $. As the evil client later explores the other path
1,3,5,6 it has to send notifications about updates. Our client can ignore
them until the node 6. That is because our client calls ``\emph{get values}''
only at the node 6. When the evil client gets to the node 6 along the path, it
has to send a notification with the node 6 and the path 1,3,5,6. Our client
reacts to the notification by recomputing its state at the dependent node 7 the
same way as before. Namely, it first issues a new ``\emph{get values}'' query at
the node 6 for both paths. Since the evil client knows now both possible values
of \texttt{a} at the node 6, its answer can be a formula $ a = 1 \vee a = 2 $.
So, our client now computes the correct interval [1,2] for \texttt{b} at the
node 7. Note that the change of the state at the node 7 may trigger
re-computations of abstract states at dependent nodes along the path 7,8.

It may further happen that the evil client stops execution of the path 1,3,5,6
before reaching the node 6. One reason for that can be a detected infeasibility
of the path, another reason can be a client's decision to spend remaining
analysis time in other parts of the program, e.g.~in the path 7,8. While the
first case implies for our client a sound value [1,1] of \texttt{b} in the
node 7, the other case implies an unsound result. It may also happen that our
client explores the whole program even before the evil client explores the path
1,3,5,6. If our client terminates without waiting for the evil one, then it can
again get an unsound result.

All these situations can be resolved if our client can ask others (through the
server) whether they have already built over-approximations of the program's
memory. It does not matter whether a client builds an over- or
under-approximation. The query can always be answered (often quite trivially).
In particular, a client may answer \emph{yes} to a query ``\emph{is memory
over-approximated?}'' only if its current abstract state space captures the
program's memory for all feasible paths.

In our example, if the evil client early terminates exploration of the path
1,3,5,6 due to its infeasibility, then its answer \emph{yes} to the query
``\emph{is memory over-approximated?}'' (together with the same answer from all
other clients) implies that our client may terminate with ``succeess'' state
(the memory is successfully over-approximated). On the other hand, if the evil
client never answers \emph{yes} to the query till the time-out, then our client
correctly terminates with ``failure'' state (i.e.~it failed to compute an
over-approximation). Finally, since our client builds an over-approximation and
it uses the query ``\emph{get values}'', it cannot terminate with the
``succeess'' state until all clients answer \emph{yes} to the query.

Waiting until the time-out with repetitive asking clients ``\emph{is memory
over-approximated?}'' is a wast of resources. Moreover, our client may terminate
with ``failure'' just because the evil client never responds \emph{yes} to the
query, e.g.~because it computes an under-approximation. Our client may solve the
issues by a simple strategy which we call ``safety outdate''. First it estimates
a time point before the time-out. Before this time point it performs its
analysis as described above. At the time point it marks for re-computation each
state in its abstract state space, which was directly or indirectly computed
from values received from other clients (via ``\emph{get values}''). And from
that time point on the client never performs ``\emph{get values}'' query. The
time point should be estimated s.t. the client is able to recompute the marked
(and dependent) parts of its state space before the time-out. In our example, at
the time point our client marks for re-computation abstract states attached to
nodes along the path 7,8 (for both paths leading to the node 7). The client then
evaluates the edge (6,7) without performing ``\emph{get values}'' query, i.e.~it
uses its interval [1,2] of \texttt{a} at the node 6. Note that the
re-computation of marked states does not imply that all information received
from other clients before the time point is lost. Typically, some invariant
properties can be preserved in strongly connected components. They can still
have a (significant) impact on the precision of a client's result.

Finally, there are situations when a client may apply the safety outdate even
before the estimated time point, for example, when all other clients compute
under-approximations or when they lose a chance to achieve over-approximations
due to failures in their analyses, e.g.~an SMT solver cannot decide
satisfiability of some formula, etc. For that purpose the protocol offers
clients a query ``\emph{can improve memory over-approximation?}''. A client
returns \emph{yes}, if it can make a progress towards memory over-approximation.
Note that the answer \emph{yes} does not imply the client will necessarily ever
do such progress. The safety outdate can be performed once all clients respond
\emph{no} to the query.

\paragraph{Summary}

The query ``\emph{get values}'' allows clients to exchange information about
memory content (via formulae). All other queries ``\emph{on location
outdated}'', ``\emph{is memory over-approximated?}'', and ``\emph{can improve
memory over-approximation?}'' are the special queries. They allow clients to
achieve soundness of their results. The final step of the integration of the
protocol into a client is a check whether there is no communication scenario,
which would produce an unsound result.

\paragraph{Boosting convergence}

We further extended the protocol by a concept allowing clients to boost
convergence of their analyses to final results.
Due to space limitation we omit its presentation here. An interested reader may
find its detailed description in our technical report~\cite{TechnicalReport}.
Note that our evaluation uses this concept.

\subsection{Canonical program}
\label{sec:CanonicalProgram}

A \emph{canonical program} is a model of the program's instruction counter. Its
purpose is to allows clients (operating on their internal program
representations) to specify the counter and program paths in queries
(like~``\emph{get values}'') uniformly.

A control-flow graph is a popular and widely used program representation in
analysers. Furthermore it also models the instruction counter. We can thus take
and adapt the code of some chosen analyser and use it to easily and quickly
implement a program utility for building canonical programs (used then by all
clients). The adaptation of the client's code might involve some pre- and/or
post-processing of a raw control-flow graph produced by the client's code. A
resulting control-flow graph (canonical program) must have the following
properties:

The instruction counter is modeled by nodes and edges represent possible
transitions of the counter. Each sub-program is modeled by a separate component
with a single entry and exit node. Each node is labeled by a set of indices of
the program's lexical tokens. We describe the computation later. We distinguish
two kinds (labels) of edges: \emph{branching} and \emph{call}. An edge is
branching if its head node has out-degree at least 2. An edge is a call edge if
the label of its head node contains indices of all tokens which correspond to a
call expression of the program.

Before we explain computation of labels of nodes we define meaning of nodes:
If the instruction counter is at a certain node of a canonical program
then it means that the instruction counter is at the position in the source code
s.t. indices of all lexical tokens of an instruction to be executed next belong
to the label of the node and indices of all lexical tokens of any predecessor
instruction belongs to the label of some predecessor of the node.

\begin{figure}[!tb]
\begin{tabular}{c}
\begin{tabular}{l}
\token{int}{1} \token{btree\_contains}{2}\token{(}{3}\token{int}{4}
\token{key}{5}\token{,}{6}\token{btree\_node}{7}\token{*}{8}
\token{node}{9}\token{)}{10}\token{\{}{11}
\\ %
\token{~~int}{12} \token{i}{13}\token{=}{14}\token{0}{15}\token{;}{16}
\\ %
\token{~~while}{17}\token{(}{18}\token{i}{19} \token{<}{20}
\token{node}{21}\token{->}{22}\token{nkeys}{23} \token{\&\&}{24}
\token{node}{25}\token{->}{26}\token{key}{27}\token{[}{28}\token{i}{29}\token{]}{30}
\token{<}{31}\token{key}{32}\token{)}{33}
\\ %
\token{~~~~++}{34}\token{i}{35}\token{;}{36}
\\ %
\token{~~if}{37}\token{(}{38}\token{i}{39} \token{<}{40}
\token{node}{41}\token{->}{42}\token{nkeys}{43} \token{\&\&}{44}
\token{node}{45}\token{->}{46}\token{key}{47}\token{[}{48}\token{i}{49}\token{]}{50}
\token{==}{51}\token{key}{52}\token{)}{53}
\\ %
\token{~~~~return}{54} \token{1}{55}\token{;}{56}
\\ %
\token{~~if}{57}\token{(}{58}
\token{node}{59}\token{->}{60}\token{child}{61}\token{[}{62}\token{i}{63}
\token{+}{64}\token{1}{65}\token{]}{66}
\token{==}{67}\token{NULL}{68}\token{)}{69}
\\ %
\token{~~~~return}{70} \token{0}{71}\token{;}{72}
\\ %
\token{~~return}{73} \token{btree\_contains}{74}\token{(}{75}\token{key}{76}\token{,}{77}
\token{node}{78}\token{->}{79}\token{child}{80}\token{[}{81}\token{i}{82}
\token{+}{83}\token{1}{84}\token{]}{85}\token{)}{86}\token{;}{87}
\\ %
\token{\}}{88}
\end{tabular}

\\
(a)
\\
\begin{tabular}{ccc}
\begin{tabular}{c}
\includegraphics[scale=0.6]{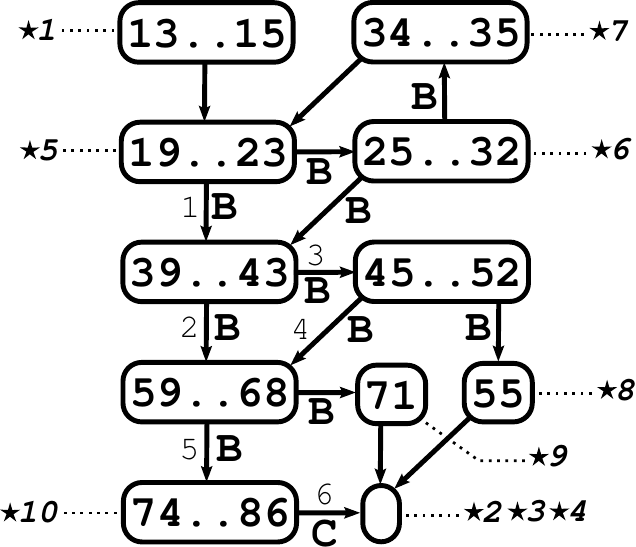}
\\
(b)
\\ \\
\begin{tabular}{c||c|c|c|c}
Identifier & \texttt{btree\_contains} & \texttt{key} & \texttt{node} & \texttt{i}
\\ \hline
Segment & 1 & 2 & 3 & 4
\end{tabular}
\\
(d)
\end{tabular}
& ~~~ &
\begin{tabular}{c}
\includegraphics[scale=0.6]{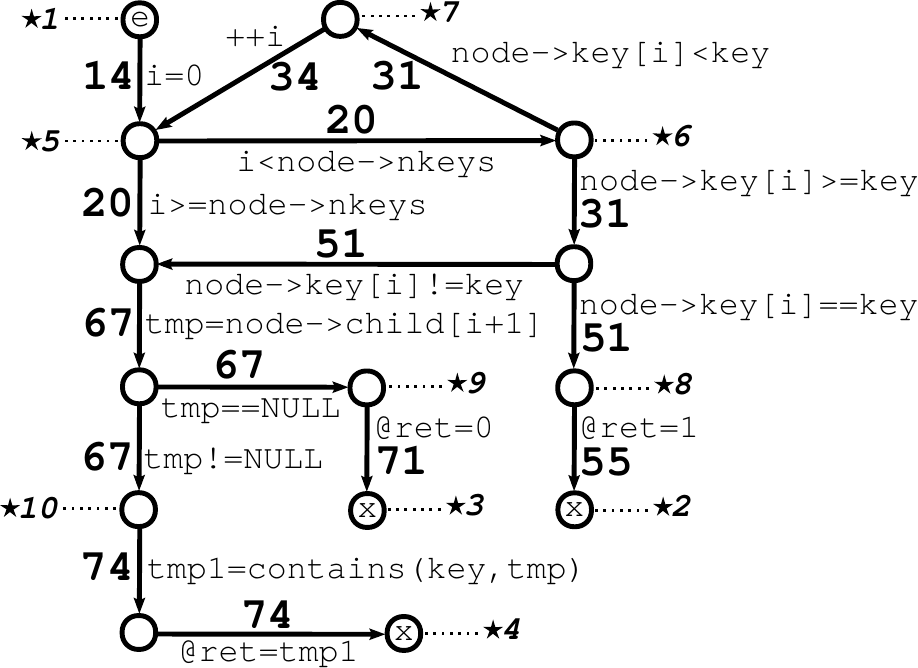}
\\
(c)
\end{tabular}
\end{tabular}
\end{tabular}
\caption{(a) A C function with lexical tokens indexed. (b) A canonical program
of the function. Labels \texttt{B} and \texttt{C} stand for branching and call
kinds respectively. Some edges are marked by numbers for purposes of
presentation. (c) An internal program representation of the function. Bold
numbers labelling edges are indices of root tokens of the corresponding C
expressions in (a). Symbols $ \star1 $, ..., $ \star10 $ represent links between
nodes of (b) and (c). (d) A mapping between (l-value) identifiers of (a) and
segments.} %
\label{fig:BTreeContains}
\end{figure}

Each edge of a canonical program corresponds to some part of program's source
code (which is executed whenever moving from its head to tail node). Therefore,
according to the meaning of nodes above we compute the label of a node as a
union of indices of lexical tokens of source code which correspond to at least
one out-going edge from the node. This construction assumes that whenever two
edges share some source code tokens, then the edges have the same head node.

Consider the C function (with indexed lexical tokens) depicted in
Fig.~\ref{fig:BTreeContains}~(a). A canonical program for this function is
depicted in Fig.~\ref{fig:BTreeContains}~(b) (for now ignore links $ \star1 $,
..., $ \star10 $). The left top node is labeled by the set $ \{13,14,15\} $.
This node thus represents the instruction counter right before execution of the
assignment \texttt{i=0}. By taking its only out-edge we get to node $ \{19..23\}
$. It represents the program counter right before execution of the condition
\texttt{i<node->nkeys} in the while loop. We can further see that indices of
some lexical tokens do not belong to label of any node, e.g.~semicolons, some
brackets, operators, keywords, etc. Their purposes are captured by the shape of
the graph. But it does not mean that we have to exclude them. It depends on our
conventions. Finally, observe that the right bottom node (the exit node) has
an empty set of labels.

Information about indices of program's lexical tokens may not be directly
available in an analyser. On the other hand information about program lines is
available practically always. We can get indices of tokens by constructing
one-to-one mapping between program lines and tokens: we only syntactically
reshape the original program s.t. each lexical token is put into a separate
line. Then we construct the canonical program from the reshaped program. Note
that for C programs there are already tools available for this
functionality~\cite{BugstURL,CPAcheckerURL}.

\subsection{Context: a specification of a set of paths in a canonical program}
\label{sec:Context}

Here we propose a simple specification for a set of program paths in a
canonical program with restricted expressivity. The restricted expressivity
is not an issue, because conversions of paths between a canonical program and an
internal program representation often loses precision (see the next section).

A \emph{filter} is a set of kinds of edges of a canonical program. So, there are
only four filters: $ \emptyset $, $ \{ \textit{branching} \} $, $ \{
\textit{call} \} $, and $ \{ \textit{branching}, \textit{call} \} $. A
\emph{context} constructed for a given filter is a list of edges whose kinds
belong to the filter. A path in a canonical program belongs to a context if and
only if the context is equal to a list of edges constructed from the path s.t.
each edge with a kind belonging to the construction filter of the context is
preserved and any other is removed. Finally, a pair ``node, context'' represents
each path which belongs to the context and which goes from the program's entry
to the node.

Let us consider the node $ \{13..15\} $ from the canonical program in
Fig.~\ref{fig:BTreeContains}~(b). If we couple it with the empty context []
constructed for the filter $ \emptyset $ then the pair represents all paths
reaching the node. If we couple it with the context [6,6,6,6] (for brevity we
use numbers marking edges) constructed for $ \{\textit{call}\} $ filter, then
the pair represents all possible paths to the node performing 4 recursive calls.

\paragraph{Multi-threading:}
\label{sec:Multi-threading}

A construct creating a new thread can be modeled in a canonical program by an
artificially introduced branching (e.g.~by two parallel edges with special
labels). Different threads can then be distinguished via a context.

\subsection{Links between canonical and internal program representation}
\label{sec:MappingCanonicalInternalProgram}

We present a simple recipe how to compute links (mapping) between nodes in a
canonical program and nodes of an internal program representation. They allow
a client to convert a pair ``node, context'' from a canonical program to an
internal program representation and back. The recipe requires that a client
builds its internal program representation from the reshaped program (see end of
Sec.~\ref{sec:CanonicalProgram}).

Let $ (h,t) $ be an edge of an internal program representation s.t. there is a
source code line associated with the edge, and let $ n $ be a node of the
canonical program containing the line in its label. We extend the mapping
between nodes in any of the following three cases: %
(1) If in-degrees of both $ n $ and $ h $ are zeros, then link $ h $ with $ n $.
(2) If $ t $ is an exit node and $ n $ has a successor with the empty label and
with the out-degree 0, then link $ t $ with the successor of $ n $. (3) If all
successor edges of $ t $ have lines associated and all the lines belong to the
label of a single successors node $ n' $ of the node $ n $ and $ n \neq n' $,
then link $ t $ with $ n' $.

Fig.~\ref{fig:BTreeContains}~(c) depicts an internal program
representation (a control-flow graph) for the C function at
Fig.~\ref{fig:BTreeContains}~(a). Links between nodes of
Fig.~\ref{fig:BTreeContains}~(b) and (c) were computed by the recipe above.
Namely, the link $ \star1 $ was set according to the case (1), links $
\star2,\star3,\star4, $ according to the case (2), and all others according to
the case (3). Observe there are nodes without links. A client cannot issue
queries at these nodes. Also, it has to answer \emph{true} for ``\emph{get
values}'' queries there. Finally, conversions of paths between internal program
representation and canonical program may lose precision. Let us consider a path
[14,20,67,67,74] in Fig.~\ref{fig:BTreeContains}~(c) (we use numeric labels for
brevity). Using links $ \star1,\star4,\star5,\star6,\star10 $, for a filter $
\{\textit{call}, \textit{branching}\} $ we can express the path by two contexts
[1,2,5,6] and [1,3,4,5,6] (we use numeric labels for brevity) s.t. we consider a
union of their paths, or for a filter $ \{\textit{call}\} $ by a context [6].
But in both cases we lose precision.

\subsection{Canonical memory}
\label{sec:CanonicalMemory}

Here we discuss how clients can uniformly encode or decode information about
memory in answers from ``\emph{get values}'' queries in order to use it in their
(different) memory representations. We call the uniform representation a
\emph{canonical memory}.

Addressing of global and stack memory-referencing identifiers is based on the
common segment-offset model: an address is simply a pair of numbers
\emph{segment} (an identifier of some memory pool) and \emph{offset} (a shift to
a certain byte inside the pool). An offset is a non-negative integer. A segment
is either 0 (a representation of the \texttt{NULL} pointer) or a unique positive
integer associated with a program identifier. We put identifiers into a list
sorted by their token indices and segments are then indices of the identifiers
in the list. In Fig.~\ref{fig:BTreeContains}~(d) we see assignments of segments
to identifiers of the function in Fig.~\ref{fig:BTreeContains}~(a). A sequence
of bytes starting at a given segment and offset can have any of the following
\emph{type} interpretations: %
(1) \texttt{i8,i16,i32,i64}/\texttt{u8,u16,u32,u64} -- a signed/unsigned
8,16,32,64-bit integer. (2) \texttt{f32,f64} -- a 32,64-bit floating point
number. (3) \texttt{seg,off} -- a representation of addresses (pointers). Data
of composed types have to be communicated per their attributes.

A \emph{dereference} is a triple consisting of a segment expression, an offset
expression, and a type. A \emph{segment expression} is either a segment, or a
dereference of the type \texttt{seg}. An \emph{offset expression} is any integer
expression possibly containing dereferences of any integer or \texttt{off} type.
A dereference represents a type-interpreted values of bytes pointed to by its
segment and offset expressions. A dereference is called \emph{basic} if neither
segment nor offset expression contains any dereference.

A value in the memory referenced by a program identifier is directly denoted by
a basic dereference. A value stored in a \emph{non-leaked} memory block in the
program's heap is always accessible (directly or indirectly) from some program
identifier. This ``access path'' from the identifier can always be expressed in
our model by nesting dereferences in segment and offset expressions. %
For example, we express a value denoted by an expression \texttt{node->nkeys}
appearing in the C function at Fig.~\ref{fig:BTreeContains}~(a). A value of the
pointer \texttt{node} we express by basic dereferences \tmb{3}{0}{seg} and
\tmb{3}{0}{off} (see Fig.~\ref{fig:BTreeContains}~(d)). If we assume that the
attribute \texttt{nkeys} is of \texttt{int} type and it has an offset 12 in the
structure \texttt{btree\_node}, then the value is represented by a non-basic
dereference \tmb{\tmb{3}{0}{seg}}{\tmb{3}{0}{off}+12}{i32} in our model. Note 
that records deeper in the call stack are accessed exactly
the same way as the memory in the program's heap.

\subsection{Client}
\label{sec:Client}

Each client has to implement the following functions (formalising queries
discussed in Sec.~\ref{sec:CommunicationQueries} in terms of
Sec.~\ref{sec:CanonicalProgram}, \ref{sec:Context}, and
\ref{sec:CanonicalMemory}):

\begin{itemize}
\item[$ \bullet~ $] \texttt{get\_values(node,context,dereferences) -> formula}\\
A client is queried for values in the memory referenced by the passed set of
\texttt{dereferences}. The pair \texttt{node,context} describes program states
where to read the values. A client's answer must be an \emph{over-approximation}
of its \emph{current} knowledge about the referenced memory. The answer must be
encoded as a quantifier-free first order logic formula over dereferences in
SMT-LIB2~\cite{SMTLIBURL} format. A formula may only contain interpreted symbols
from theories of integers, Peano arithmetic, and reals. Dereferences are the
only allowed uninterpreted symbols. We encode a dereference $ (s,o,t) $ as an
application of a binary function symbol $ \texttt{DEREF\_}t $ to arguments $ s $
and $ o $. %
\item[$ \bullet~ $] \texttt{on\_location\_outdated(node,context) -> nothing}\\ %
A client is notified that the abstract state space of some client has been
updated. The pair \texttt{node,context} identifies program states the update was
relevant for. %
\item[$ \bullet~ $] \texttt{is\_memory\_over\_approximated() -> bool} \\ %
A client may answer \texttt{true} only if its current abstract state space
captures the program's memory for all feasible paths. In particular, the client
is supposed to return \texttt{false} in any of the following cases:
\begin{enumerate}
\item Its function \texttt{can\_improve\_memory\_over\_approximation} returns
\texttt{true}. %
\item It has not sent a notification \texttt{on\_location\_outdated} to the
server yet about an update of some of its abstract state. %
\item An over-approximation of the program's memory cannot be achieved due to
failures in its computation, e.g.~failures in an SMT solver, etc.
\end{enumerate}
\item[$ \bullet~ $] \texttt{can\_improve\_memory\_over\_approximation() ->
bool}\\ %
A client returns \texttt{true} if it can make a progress towards memory
over-approximation. The answer \texttt{true} implies only a possibility of the
progress.
\end{itemize}

\subsection{Server}
\label{sec:Server}

A server has to provide the following functions to clients (expressed in terms
of Sec.~\ref{sec:CanonicalProgram}, \ref{sec:Context}, and
\ref{sec:CanonicalMemory}):

\begin{itemize}
\item[$ \bullet $] \texttt{get\_values(node,context,dereferences) -> formula}\\%
The server broadcast the query to functions \texttt{get\_values} of all clients.
Then it returns a \emph{conjunction} of received formulae to the client issuing
the query. %
\item[$ \bullet $] \texttt{on\_location\_outdated(node,context) -> nothing}\\ %
The server broadcast the notification to functions of the same name of all
clients. Note that each client \emph{has to} call this function whenever its
abstract state space is changed. %
\item[$ \bullet $] \texttt{is\_memory\_over\_approximated() -> bool}\\ %
The server broadcast the query to functions of the same name of all clients.
Then it returns the \emph{conjunction} of individual answers. %
\item[$ \bullet $] \texttt{can\_improve\_memory\_over\_approximation() ->
bool}\\%
The server broadcast the query to functions of the same name of all clients.
Then it returns the \emph{disjunction} of individual answers.
\end{itemize}

%

%
%

\section{Evaluation: Box, Polka, and Symbolic execution}
\label{sec:Evaluation}

We wanted to know how much clients may ideally improve their results using our
communication model. We thus experimentally evaluated a limit case, where
clients were offered maximum opportunities for the communication: a client can
communicate at each node where a loss of precision may occur (e.g.~joins, loop heads, nodes around pointer or other (unsupported) operations), and there is a
negligible overhead of message delivery.

We embedded three clients into a single tool~\cite{StatorURL}: two abstract
interpreters Box (intervals)~\cite{ApronURL}, Polka
(polyhedrons)~\cite{ApronURL}, and the classic Symbolic
execution~\cite{SE_King76}. The tool implements the server. Clients share a
single internal program representation modelling also a canonical program. Since
all clients run on a single main thread, they perform computations in small
regularly interleaved steps. A step corresponds to an update of an abstract
state space by taking one or more edges which all always share either head or
tail node. A client determines by itself which edges it will process in what
computation step. A client may issue communication queries to the server only
during its step. Responses from other clients are also computed and returned in 
that step.

We evaluated clients in five different configurations. Each configuration
specifies what clients are used and whether they can communicate or not. We
denote configurations using the following abbreviations: b*p*s, b+p+s, b+p, b+s,
and p+s. Symbols `b', `p', and `s' stand for Box, Polka, and Symbolic execution
respectively, and `+' and `*' stand for communication enabled and disabled
respectively. For each configuration we assume that either all clients
communicate with each other (the use of `+') or none of them (the use of `*').

We performed the evaluation on the SV-COMP~2015~\cite{SVCOMPURL} benchmark
suite, revision 571. In order to make the evaluation manageable for us, we put a
requirement that the whole evaluation (all five configurations) should finish
within one week of continuous computation\footnote{One process on a server:
2xIntel Xenon E5-2650 @ 2GHz, 64GB RAM, Debian~4.6.3.}. We thus picked 10
randomly chosen benchmarks from each directory (or less if there was not enough)
and so we got 473 benchmarks. We further set a time-out at 2.5 minutes and a
memory-out 512MB for each client in each configuration. It means, for example,
that b+p had the time-out 5 minutes and the memory-out 1024MB, b*p*s had the
time-out 7.5 minutes and the memory-out 1536MB, etc. Remember, clients share
time (steps are interleaved) and memory (all run on a single thread) within a
configuration.

We compared results of each combination of configurations. The comparison was
always done per client: given two configurations and a client appearing in both
of them we compare only results of that client (i.e. we ignore results from all others). Note that clients are independent, so they produce independent outputs.


\begin{figure}[!htb]
\begin{center}
\begin{tabular}{c}
\begin{tabular}{|cc||ccccc|ccccccc|}
\hline
\multicolumn{2}{|c||}{Configuration} &
\multicolumn{5}{c|}{Comparison per node} &
\multicolumn{7}{c|}{Comparison per benchmark}
\\
1st & 2nd &
fail & neq & eq & 1st & 2nd &
fail & neq & 1st & 2nd & 1st! & 2nd! & eq!
\\ \hline
b*p*s & ~b+p+s &
13 & 246 & 13021 & 627 & 9563 & 
1 & 20 & 56 & 191 & 23 & 148 & 35
\\
b*p*s & b+p &
13 & 342 & 15042 & 613 & 10043 & 
1 & 37 & 46 & 243 & 7 & 186 & 34
\\
b*p*s & b+s &
13 & 309 & 12444 & 450 & 8202 & 
1 & 10 & 36 & 128 & 28 & 115 & 71
\\
b+p+s & b+p &
13 & 1167 & 25025 & 3707 & 1851 & 
1 & 42 & 120 & 123 & 43 & 48 & 97
\\
b+p+s & b+s &
13 & 142 & 24753 & 1170 & 708 & 
1 & 14 & 66 & 34 & 43 & 17 & 155
\\
b+p & b+s &
13 & 889 & 21311 & 1798 & 2899 & 
1 & 41 & 106 & 76 & 56 & 28 & 99
\\
\hline
\end{tabular}

\\ \\
\begin{tabular}{|cc||ccccc|ccccccc|}
\hline
\multicolumn{2}{|c||}{Configuration} &
\multicolumn{5}{c|}{Comparison per node} &
\multicolumn{7}{c|}{Comparison per benchmark}
\\
1st & 2nd &
fail & neq & eq & 1st & 2nd &
fail & neq & 1st & 2nd & 1st! & 2nd! & eq!
\\ \hline
b*p*s & ~b+p+s &
20 & 487 & 10616 & 401 & 12672 & 
1 & 42 & 38 & 199 & 16 & 157 & 43
\\
b*p*s & b+p &
26 & 319 & 11691 & 168 & 12923 & 
1 & 39 & 23 & 224 & 5 & 183 & 51
\\
b*p*s & p+s &
20 & 609 & 10689 & 455 & 12119 & 
1 & 57 & 44 & 196 & 12 & 136 & 50
\\
b+p+s & b+p &
14 & 663 & 30462 & 2217 & 1489 & 
1 & 29 & 93 & 87 & 50 & 53 & 137
\\
b+p+s & p+s &
0 & 253 & 32268 & 1307 & 446 & 
0 & 19 & 74 & 40 & 53 & 24 & 182
\\
b+p & p+s & 
14 & 691 & 30187 & 1960 & 1903 & 
1 & 46 & 115 & 88 & 58 & 33 & 129
\\
\hline
\end{tabular}

\end{tabular}
\end{center}
\caption{Comparison of invariants for clients Box (top) and Polka (bottom).
Meaning of columns from left: ``1st''/``2nd'' - 1st/2nd compared configuration,
``fail'' - failures of Z3, ``neq'' - incomparable (neither is stronger), ``eq''
- logically equal, ``1st''/``2nd'' - 1st/2nd configuration has stronger
invariant ; ``fail'' - at least one Z3 failure, ``neq'' - contains incomparable
invariants, ``1st''/``2nd'' - 1st/2nd has at least one stronger invariant than
2nd/1st configuration, ``1st!''/``2nd!'' - at least one stronger but no weaker
invariant in 1st/2nd than in 2st/1nd configuration, ``eq!'' - all invariant are
logically equal.} \label{fig:Invariants}
\end{figure}

We focused on two kinds of measurements. First, we compared the precision of
invariants computed by clients Box and Polka. Symbolic execution does not
provide this kind of information. The clients attempt to compute for each node a
strongest invariant over-approximating all concrete states which can be seen at
the node. The results are presented in Fig.~\ref{fig:Invariants}. The numbers
for ``Comparison per node'' are summary counts of nodes of all considered
benchmarks together. And the numbers for ``Comparison per benchmark'' are simply
counts of benchmarks. Note that for each client there were only considered those
benchmarks for which the client terminated with the state ``Success'' in both
compared configurations. We can observe the following facts about the data in
Fig.~\ref{fig:Invariants}:
\begin{itemize}
\item \emph{Each configuration may bring us improvements over others}: We can
clearly see this phenomena for all pairs of configurations in both kinds of
comparisons in tables of both clients. %
\item \emph{There is no configuration strictly dominating all others}: We can
only read patterns in the data, like:
\begin{itemize}
\item A configurations with communicating clients gives us at least one order of
magnitude more precise invariants than isolated clients. %
\item More communicating clients, more strengthened invariants. %
\item Count of incomparable invariants and lower count of strengthened
invariants can be expected in the same order of magnitude. %
\item More improved invariants typically yields more improved benchmarks,
i.e.~improvements are rather regularly distributed than highly concentrated in
few benchmarks. Nevertheless, a degree of correlation is sensitive to kinds of
clients appearing in configurations, cf.~fourth and sixth rows for both Box and
Polka. %
\end{itemize}
Observations made for invariants can easily be adopted to similar
observations for benchmarks.
\end{itemize}

Data in both tables in Fig.~\ref{fig:Invariants} for the configuration b*p*s
show that the communication also weakened some invariants. Since this might be
counter-intuitive, we show on a simple example how a precise information
delivered to an analysis may actually lead to a worse result: consider the
interval analysis with a widening applied on a C code ``... \texttt{while (i<10)
++i;} ...'' s.t. it reaches the \texttt{while} statement with the value $
\texttt{i}=[0,50] $. This implies the result $ [0,50] $ for \texttt{i} at the
loop head. If we deliver a precise value $ [0,0] $ to \texttt{i} before the
loop, then the widening operator will produce a weaker result $ [0,\infty] $ at
the loop head.

\begin{figure}[!htb]
\begin{center}
\begin{tabular}{c}
\begin{tabular}{|cc||ccc|ccc|ccc|ccc|}
\hline
\multicolumn{2}{|c||}{Configuration} &
\multicolumn{3}{c|}{Success} &
\multicolumn{3}{c|}{Time-out} &
\multicolumn{3}{c|}{Memory-out} &
\multicolumn{3}{c|}{Crash}
\\
1st & 2nd &
eq & 1st & 2nd &
eq & 1st & 2nd &
eq & 1st & 2nd &
eq & 1st & 2nd
\\ \hline
b*p*s & ~b+p+s &
251 & 35 & 24 & 
10 & 0 & 38 & 
137 & 19 & 0 & 
7 & 14 & 6
\\
b*p*s & b+p &
285 & 1 & 25 & 
10 & 0 & 3 & 
143 & 13 & 0 & 
7 & 14 & 0
\\
b*p*s & b+s &
230 & 56 & 23 & 
10 & 0 & 200 & 
0 & 156 & 0 & 
7 & 14 & 3
\\
b+p+s & b+p &
275 & 0 & 35 & 
13 & 35 & 0 & 
137 & 0 & 6 & 
7 & 6 & 0
\\
b+p+s & b+s &
241 & 34 & 12 & 
45 & 3 & 165 & 
0 & 137 & 0 & 
10 & 3 & 0
\\
b+p & b+s &
245 & 65 & 8 & 
12 & 1 & 198 & 
0 & 143 & 0 & 
7 & 0 & 3
\\
\hline
\end{tabular}

\\ \\
\begin{tabular}{|cc||ccc|ccc|ccc|ccc|}
\hline
\multicolumn{2}{|c||}{Configuration} &
\multicolumn{3}{c|}{Success} &
\multicolumn{3}{c|}{Time-out} &
\multicolumn{3}{c|}{Memory-out} &
\multicolumn{3}{c|}{Crash}
\\
1st & 2nd &
eq & 1st & 2nd &
eq & 1st & 2nd &
eq & 1st & 2nd &
eq & 1st & 2nd
\\ \hline
b*p*s & ~b+p+s &
263 & 21 & 30 & 
10 & 0 & 20 & 
137 & 21 & 0 & 
7 & 14 & 6
\\
b*p*s & b+p &
283 & 1 & 28 & 
10 & 0 & 2 & 
143 & 15 & 0 & 
7 & 14 & 0
\\
b*p*s & p+s &
267 & 17 & 27 & 
10 & 0 & 15 & 
143 & 15 & 0 & 
7 & 14 & 4
\\
b+p+s & b+p &
289 & 4 & 22 & 
10 & 20 & 2 & 
137 & 0 & 6 & 
7 & 6 & 0
\\
b+p+s & p+s &
287 & 6 & 7 & 
21 & 9 & 4 & 
137 & 0 & 6 & 
11 & 2 & 0
\\
b+p & p+s &
294 & 17 & 0 & 
12 & 0 & 13 & 
143 & 0 & 0 & 
7 & 0 & 4
\\
\hline
\end{tabular}

\\ \\
\begin{tabular}{|cc||ccc|ccc|ccc|ccc|}
\hline
\multicolumn{2}{|c||}{Configuration} &
\multicolumn{3}{c|}{Success} &
\multicolumn{3}{c|}{Time-out} &
\multicolumn{3}{c|}{Memory-out} &
\multicolumn{3}{c|}{Crash}
\\
1st & 2nd &
eq & 1st & 2nd &
eq & 1st & 2nd &
eq & 1st & 2nd &
eq & 1st & 2nd
\\ \hline
b*p*s & ~b+p+s & 
151 & 3 & 0 & 
155 & 0 & 35 & 
119 & 16 & 0 & 
12 & 17 & 1
\\
b*p*s & b+s & 
148 & 6 & 7 & 
155 & 0 & 153 & 
0 & 136 & 0 & 
10 & 19 & 0
\\
b*p*s & p+s & 
149 & 5 & 0 & 
155 & 0 & 34 & 
124 & 11 & 0 & 
11 & 18 & 0
\\
b+p+s & b+s & 
147 & 4 & 8 & 
190 & 0 & 118 & 
0 & 119 & 0 & 
10 & 3 & 0
\\
b+p+s & p+s & 
148 & 3 & 1 & 
185 & 5 & 4 & 
119 & 0 & 5 & 
11 & 2 & 0
\\
b+s & p+s & 
148 & 7 & 1 & 
189 & 119 & 0 & 
0 & 0 & 124 & 
10 & 0 & 1
\\
\hline
\end{tabular}

\end{tabular}
\end{center}
\caption{Comparison of termination states for clients Box (top), Polka (middle),
and Symbolic execution (bottom). Columns of ``Configuration'': ``1st''/``2nd'' -
1st/2nd compared configuration~;~All other columns: ``eq'' - equal state,
``1st''/``2nd'' - 1st/2nd configuration has the state while 2nd/1st has some
other.} \label{fig:Termination}
\end{figure}

In the second measurement we focused on comparison of termination states of
individual clients as they are used in different configurations. We distinguish
termination states ``Success'', ``Time-out'', ``Memory-out'', and ``Crash'', all
with obvious meanings. Results are presented in Fig.~\ref{fig:Termination}.
Numbers in each table represent counts of benchmarks. We can observe the
following facts in data in Fig.~\ref{fig:Termination}:
\begin{itemize}
\item \emph{Consumption of resources via communication does not imply a decrease
of successful termination}: Considering ``Success'' data for all configurations
comparing with b*p*s for all clients, the communication caused a loss of success
termination states in the following percentages:

\begin{tabular}{cc||c|c|c}
1st & 2nd & Box & Polka & Sym.exec.
\\ \hline
b*p*s & ~b+p+s &
3.8 &  
-3.2 & 
1.9    
\\
b*p*s & b+p &
-8.4 & 
-9.5 & 
-
\end{tabular}
~~~~~
\begin{tabular}{cc||c|c|c}
1st & 2nd & Box & Polka & Sym.exec.
\\ \hline
b*p*s & ~b+s &
11.5 &  
- &
-0.6    
\\
b*p*s & ~p+s &
- &
-3.5 & 
3.2   
\end{tabular}

In 5 of 9 cases we see an increase of ``Success'' termination states. The
average of these numbers is -0.53\%. We may thus expect about 0.5\% increase of
``Success'' termination states on average per client due to reduced overall time
and memory consumption. 
\item \emph{Resources consumption via communication heavily depends on kinds of
clients}: This statement is based on the following patterns which dominate
data:
\begin{itemize}
\item Symbolic execution is a major source of ``Time-out'' termination states.
We can see this in the tables of all clients: Whenever the client is present,
there is a high count of time-outs. %
\item Polka is a major source of ``Memory-out'' termination states. We can see
this in the tables of all clients: Whenever the client is present, there is a
high count of memory-outs. %
\end{itemize}
\end{itemize}



We observed two kinds of crashes during the evaluation. The wast majority of
them occurs inside the Apron~\cite{ApronURL} library and the remaining crashes
occur when parsing complex initialiser lists. Although all the crashes can be
caused by our wrong use of those modules, we was unable to find the causes in a
reasonable time. Nevertheless, we can easily compute from the numbers in
Fig.~\ref{fig:Termination} that a configuration crashed on 57.2 from all 473
benchmarks on average. It means that each configuration was evaluated on 415,8
benchmarks on average without a crash. We believe the count of 415 benchmarks
still represent an evaluation of a sufficient size.

We finish this section by presenting interesting data related to the source code
of our implementation. In the table bellow we show for all tree clients numbers
of source code lines required for implementation of individual protocol
functions. The abbreviation SE stands for the Symbolic execution and the numbers
in the brackets represent a code performing pure conversion of an abstract state
to a formula. Note that the number 0 for SE indicates that this client does not
use knowledge from other client; it only provides knowledge to others.
\begin{center}
\begin{tabular}{r||c|c|c}
client's protocol function~ & ~Box~ & ~Polka~ & ~SE~\\
\hline\hline
\texttt{get\_values}~ & ~30 (+27)~ & ~29 (+76)~ & ~7 (+139) \\ 
\texttt{on\_location\_outdated}~ & 63 & 63 & 0\\ 
\texttt{is\_memory\_over\_approximated}~ & 3 & 3 & 1 \\
\texttt{can\_improve\_memory\_over\_approximation}~ & 3 & 3 & 1 \\
\hline
summary~ & 126 & 174 & 148 
\end{tabular}
\end{center}
We see that implementation of protocol functions is indeed small, about 150
lines on average. Note that queries to the server consist of few lines per
client. Namely, Box and Polka calls the server (all kinds of queries together)
on 6 lines each and the Symbolic execution on 3 lines. %
%
We encourage a reader to inspect all this communication-related source code to
see also its simplicity. A ZIP package with sources and Linux binaries together
with the evaluation results is freely available in~\cite{StatorURL}.
Installation and use are both extremely simple. Nevertheless, a reader may find
details in Appendix~B of our technical report~\cite{TechnicalReport}.

\section{Related Work}
\label{sec:RelatedWork}

There is a broad class of approaches dedicated to combining lattice-based
analyses. They are based on either direct or reduced
product~\cite{DirectReducedProduct}. The direct product is fully automatic, but
composed analyses do not interact. The reduced product is based on
(non-computable) concretisation functions. This is solved either by focusing on
particular kinds of lattices~\cite{LogicalProduct,ProductShapes} or by
an approximation~\cite{GrangerProduct}. An open product~\cite{OpenProduct}
substantially improves~\cite{GrangerProduct}, since it
removes dependencies between analyses. The only common property is a priory
given set of queries. This requirement was later relaxed
in~\cite{InteractingPlugins} by replacing the set of queries by a language of
the first order logic. Operations of all analyses are then parametrised by any
formula of the language. Composition of configurable program
analyses~\cite{CPA,CPAplus} is based on the direct product, whose precision can
be improved via relations ``transfer'', ``merge'', ``stop'', ``prec'',
``compare'', and ``strengthen''. They are defined over domains of all composed
analyses. We thus have to implement them for each combination of analyses.
Individual analyses do not have to be changed, if they share the same internal
program representation. Execution of analyses is
synchronised using a special lattice-based ``location analysis'' which is
supposed to appear among ``regular'' analyses in a combination. It defines
an exploration direction, e.g.~forward, backward. An advanced combination of
lattice-based analyses can be found in~\cite{Astree07}. It is based
on the idea of the open product with several extensions. The set of fixed
queries was replaced by an extensible set of kinds of constraints. An extension
of the set by a new kind implies extensions of only those analyses which want to
use constraints of that kind. Analyses may exchange messages through input and
output channels. Messages are elements of a separate abstract domain. They are
not always exchanged freely between analyses. An order of analyses in a
computational step matters. Typically, an analysis may freely communicate with
any predecessor. Analyses are synchronised and they share the same internal
program representation.

Approaches based on the open
product~\cite{OpenProduct,InteractingPlugins,Astree07} are closest to our model
because of independence of combining analyses. We can also find similarities
with~\cite{Astree07,InteractingPlugins} in formula-based communication. We
further share the goal to maximally reuse existing analyses
with~\cite{Astree07,CPA,CPAplus}. On the other hand, our approach allows
individual analyses to operate on different internal program representations,
analysers are extended once for all combinations, and there is no
synchronisation in computational steps (e.g.~selection of program transitions)
between clients. Finally, an obvious difference is that our approach allows to
combine other than only lattice-based analyses.

There is another broad class of approaches based on combining program analyses.
Typically, two or more particular analyses are considered, e.g.~predicate
abstraction with dynamic test generation~\cite{SMASH}, static checking and
testing~\cite{CheckNCrash,StaticDynamic}, different testing
techniques~\cite{CombiningTesting}, symbolic and concrete
execution~\cite{Concolic}, static and dynamic analyses via program
partitioning~\cite{ProgramPartitioning}, data-flow with predicate
lattices~\cite{DataflowPredicates}, pointer and numerical
analyses~\cite{ArrayBoundChecking,ValueAnalysis}, data-flow analyses in a
compiler~\cite{CombiningOptimisations,CompilerTranformations}, etc., and a
result is a new program analysis with advantages of individual analyses.
Clearly, all these analyses represent instances of the approach mentioned at the
beginning of the introduction. Our approach represents an alternative: we do not
build a new analysis from given analyses; we propose an analysis-independent
communication protocol.

\section{Discussion}
\label{sec:Discussion}

Whenever researchers have attempted to combine some program analyses so far,
they always focussed on solving a question ``when exchange what information
between combined analyses'' in order to get the best result from them. It means
that the output from their endeavour always was an \emph{algorithm} standing
above combined analyses fully specifying what will be exchanged and when.
There is no doubt this process can yield efficient algorithms (see the previous section). 
Nevertheless, nobody so far has
answered the question ``what are \emph{natural} dispositions of analyses for
their mutual cooperation'', i.e.~how well they can perform without any sort of
supervision by an algorithm standing above them. This work is supposed to fill
this gap.

From the theoretical point of view it is interesting to know, what analyses
naturally cooperate well (i.e. without any supervision) with what others. We
have already given the answer in Section~\ref{sec:Evaluation} for two popular
abstract domains of abstract interpretation (intervals and polyhedrons) and
King's symbolic execution. Another interesting question is, how much know
algorithms combining analyses actually improve over the presented natural (i.e.
unsupervised, spontaneous, unorganised) cooperation. It is, of course, expected
that such algorithms should perform better, but the question is how much.

The proposed approach can also be useful from the practical point of view. For
instance, it can aid in the process of inventing new algorithms combining
existing analyses. This process typically starts by choosing ``right''
analyses to combine. The choice is usually based on researcher's intuition. If
we realise that for $ n $ available analyses there is an exponential number ($
2^n - n - 1 $) of possible combinations, the intuition may easily lead him to a
sub-optimal choice and ad hoc trying more combinations may cost him months. Using our approach the researcher may quickly experimentally
evaluate several combinations of analyses and use the received data to improve
his intuition. The data can indeed be obtained quickly. According to results in
the Section~\ref{sec:Evaluation} he can roughly expect 150 source code lines per
analyser for integration of the protocol. Analysers with the protocol integrated
are ready to use in any possible combination. The obtained data can
also be useful in later stages of the process. For example, the researcher may
search in the data for cases where the communication produced interesting
results (impressive or pure). Their analysis may help the researcher to uncover
key principles for an efficient algorithm combining the chosen analyses.

\section{Conclusion}
\label{sec:Conclusion}

We presented an light-weight approach allowing cooperation of analysers during 
their simultaneous analysis of a given program. It suffices to integrate the
introduced communication protocol into $ n $ analysers, and we can then
immediately try any of their $ 2^n - n - 1 $ possible combinations. An analyser
communicates with others according its actual personal needs. So, there is no
central algorithm standing above analysers which would tell them when to
communicate what information. Our experimental evaluation provides an empirical
evidence that program analyses based on abstract interpretation and symbolic
execution have natural dispositions for mutual cooperation in the presented
communication model.

\bibliographystyle{plain}
\bibliography{communication_ARXIV_RP}

\end{document}